\documentclass[iop]{emulateapj}
\usepackage{amsmath}
\usepackage{hyperref}

\shorttitle{Pre-LHB Evolution of the Earth's Obliquity}
\shortauthors{G. Li and K. Batygin }

\begin{document}
\newcommand{\icarus}{Icarus}

\title{Pre-LHB Evolution of the Earth's Obliquity}
\author{Gongjie Li\altaffilmark{1} and Konstantin Batygin\altaffilmark{2}}
\affil{1. Harvard-Smithsonian Center for Astrophysics}
\affil{2. Division of Geological and Planetary Sciences, California Institute of Technology, 1200 E. California Blvd., Pasadena, CA 91125}

\email{gli@cfa.harvard.edu}


\altaffiltext{1}{1Harvard-Smithsonian Center for Astrophysics, The Institute for Theory and
Computation, 60 Garden Street, Cambridge, MA 02138, USA}


\begin{abstract}
The Earth's obliquity is stabilized by the Moon, which facilitates a rapid precession of the Earth's spin-axis, de-tuning the system away from resonance with orbital modulation. It is however, likely that the architecture of the Solar System underwent a dynamical instability-driven transformation, where the primordial configuration was more compact. Hence, the characteristic frequencies associated with orbital perturbations were likely faster in the past, potentially allowing for secular resonant encounters. In this work we examine if at any point in the Earth's evolutionary history, the obliquity varied significantly. Our calculations suggest that even though the orbital perturbations were different, the system nevertheless avoided resonant encounters throughout its evolution. This indicates that the Earth obtained its current obliquity during the formation of the Moon. 
\end{abstract}


\section{Introduction}
Obliquity variation plays a major role in the modulation of climate, as it determines the latitudinal distribution of solar radiation. For instance, according to the Milankovitch theory, the ice ages on the Earth are closely associated with the variation in insolation at high latitudes, which depends on the orbital eccentricity and orientation of the spin axis \citep[e.g.][]{Weertman76, Hays76, Imbrie82}. 

The spin-axis dynamics of the Earth-Moon system has been extensively studied in the literature and is generally well understood. At present, the obliquity variation of the Earth is regular and only undergoes small oscillations between $22.1^\circ$ and $24.5^\circ$ with a 41000 year period \citep[e.g.][]{Vernekar72, Laskar93b}. Without the Moon, the obliquity of the hypothetical Earth is chaotic, but is constrained between $0 - 45^\circ$ over billion year timescales \citep{Laskar93a, Lissauer12, Li14}. 

The difference between obliquity cycles exhibited by a Moon-less Earth and that corresponding to the real Earth arise largely as a consequence of the underlying resonant structure \citep{Laskar96}. Specifically, the spin-axis of the Earth may exhibit complex behavior if its precession resonates with the secular evolution of the Earth's orbit. The former is dominantly controlled by Solar and Lunar torques, whereas the latter is forced by long-period planet-planet interactions. In absence of the Moon, the Earth would indeed find itself residing within a multi-resonant domain signaling chaotic motion \citep{Chirikov79, Laskar93a}. The introduction of the Moon, however, accelerates the precession of the spin-axes, and detunes the system away from resonance yielding quasi-periodic evolution \citep{deSurgy97}. 

\begin{figure}[t]
\includegraphics[width=\columnwidth]{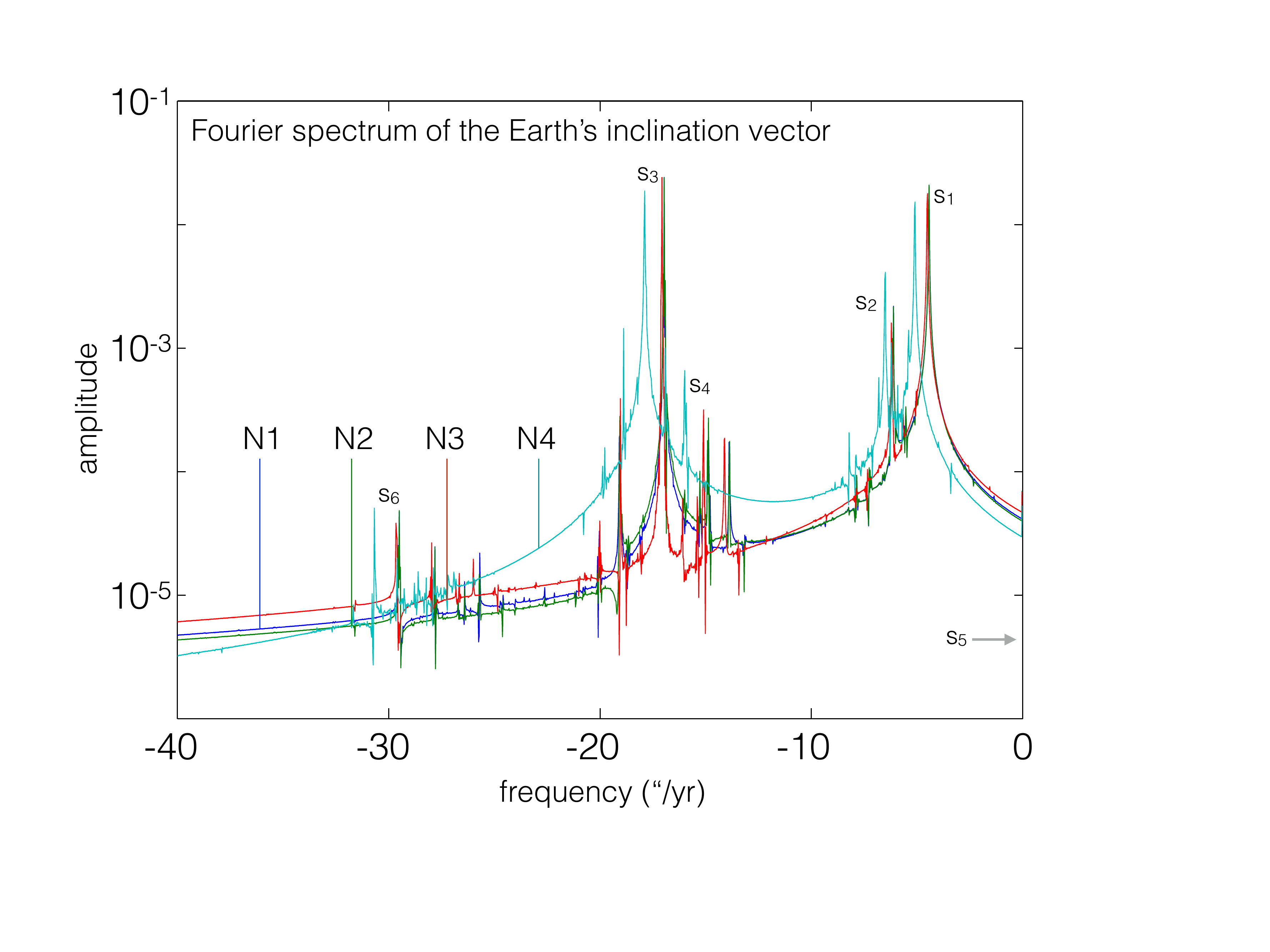}
\caption{\label{f:FT} The Fourier spectrum of the Earth's orbital parameter ($i_{\oplus} e^{i\Omega_{\oplus} t}$) obtained from an N-body simulation using the \textit{mercury6} program. The initial conditions are those identified in \citet{Batygin10}, which are compatible with the Nice model. Note that the maximum high-amplitude secular frequencies corresponding to multi-resonant conditions are $\sim -18^{''}/yr$, which is similar to the current maximum secular frequency $s_3=-18.8512^{''}/yr$.}
\vspace{0.1cm}
\end{figure}

The aforementioned real vs Moon-less Earth discussion leaves open the question of how the dynamical state of the spin-axis may have responded to changes in the orbital architecture. After all, if the Earth's orbital evolution was once characterized by more rapid secular evolution, past resonant behavior of the spin-axis cannot be ruled out a-priori. Indeed for the case of Mars, the study of \citet{Brasser11} has shown that the orbital rearrangement of the Solar System has led to a qualitative change in the dynamical behavior of the spin-axis. Hence, it is possible that in the history of the Earth-Moon system, the obliquity variation of the Earth was once significant. Correspondingly, understanding the past variation of the obliquity shines light on how the Earth obtained its current spin-orbit misalignment.  

Substantial progress has been made towards the characterization of the early dynamical evolution of the Solar System through the development of the Nice model. Qualitatively, the picture envisioned within the context of the Nice model is one where the giant planets start out on a compact orbital configuration and following a transient instability scatter onto their current orbits \citep{Tsiganis05, Levison11, Nesvorny11, Batygin12, NesvornyMorby12}. The numerous successes of the Nice model include a replication of the dynamical architecture of the outer Solar System \citep{Morby09}, the inner solar system \citep{Brasser09, Agnor11}, the formation of the Kuiper belt \citep{Levison08, BatyginBrownFraser11}, the chaotic capture of Jupiter and Neptune trojan populations \citep{Morby05, Nesvorny07} as well as its role as a trigger mechanism for late heavy bombardment (LHB) \citep{Gomes05}. 



Although the primordial state of the Solar System is not well constrained, it is likely that the giant planets resided in a multi-resonant configuration (i.e. a condition where each planet resides in mean motion resonances with each of its neighbors) at the time of nebular dispersion, as such architectures are natural outcomes of disk-driven migration \citep{Masset01, Morby07}. Under this assumption, a limited number of configurations compatible with the Nice model have been identified \citep{Batygin10, Nesvorny12}. Accordingly, in this study we extend the quantification of the Nice model by exploring the spin-axis dynamics of the Earth-Moon system within the context of pre-instability orbital configurations. 

The plan of this paper is as follows. In \textsection{2}, we analyze resonant conditions, and in \textsection{3}, we study the obliquity variation using numerical simulations. We summarize and discuss our results in \textsection{4}.

\begin{figure}[t]
\includegraphics[width=\columnwidth]{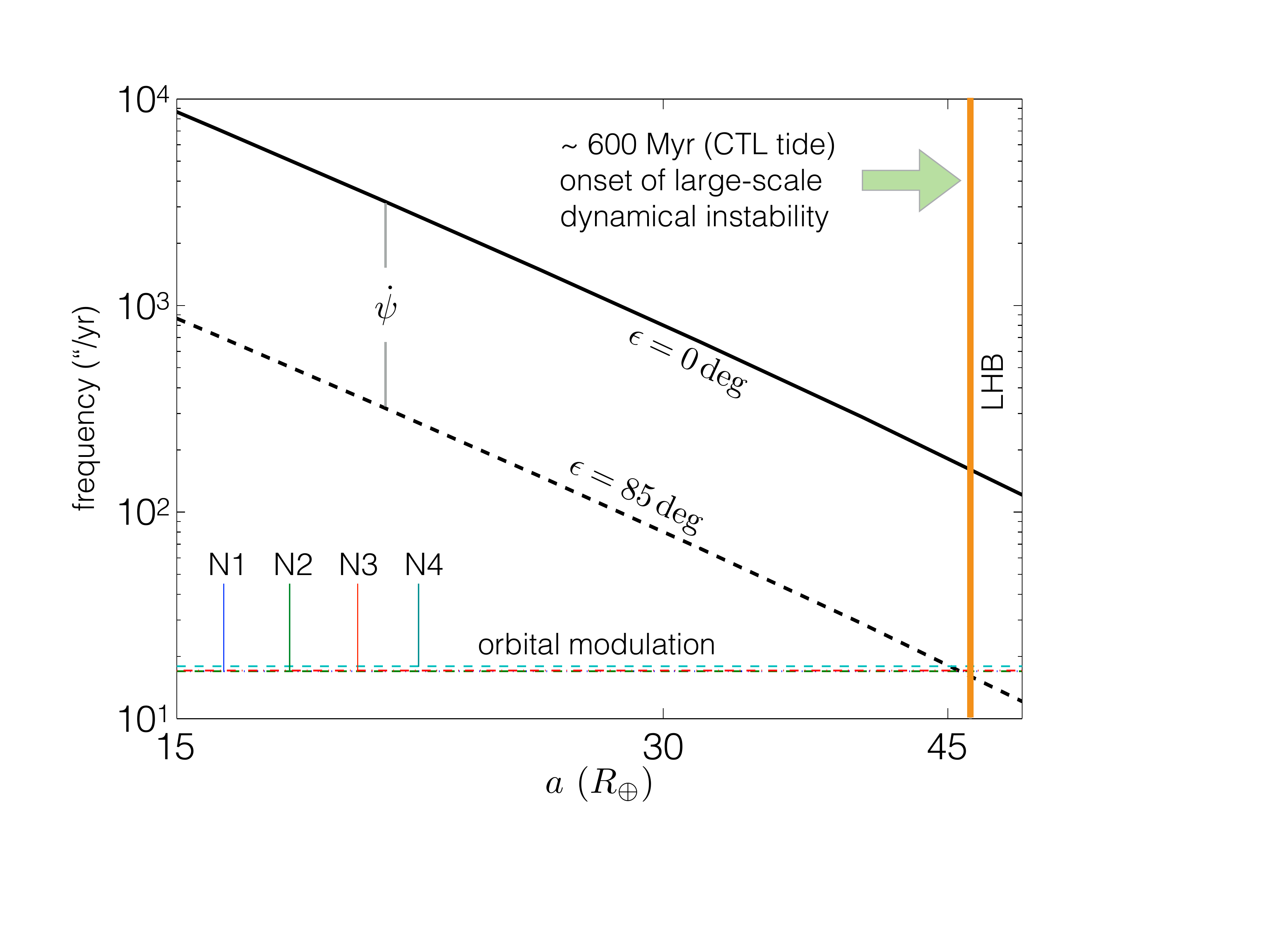}
\caption{\label{f:alphadecay} The comparison of the maximal secular frequencies of the Earth corresponding to various outer Solar System architectures and the forced spin-axis precession rate of the Earth. The presented analysis shows that the precession frequency is much bigger than the orbital frequency at low obliquity, and this indicates that there are no significant obliquity variations in the history of the Earth-Moon system due to resonances.}
\vspace{0.1cm}
\end{figure} 

\section{Spectral Analysis}
As already mentioned above, the Sun and the Moon torque the spin axis of the Earth, and the other planets in the solar system perturb the orbit of the Earth. When the two effects share the same frequencies, resonances arise and the spin-orbit angle (obliquity) undergoes large amplitude variations \citep{Colombo66, Ward73, Henrard87}. Furthermore, if the resonances overlap, the obliquity variation becomes chaotic \citep{Chirikov79, Laskar93a}. Therefore, the obliquity variation is sensitive to the two sets of frequencies. Accordingly, in this section, we investigate whether resonant motion was plausible at any point in the system's evolutionary history.

In order to obtain the dominant secular frequencies of the Earth's orbital inclination vector, we performed N-body integrations of the multi-resonant conditions identified by \citet{Batygin10} using the \textit{mercury6} orbital integration software package \citep{Chambers99}. The specific multi-resonant states onto which the giant planets were initialized are delineated in Table (\ref{t:mres}). The rows labeled N1--N4 correspond to different multi-resonant states, while the columns depict neighboring period ratios\footnote{Initial conditions where Jupiter and Saturn are locked in a second-order 5:3 resonance were not considered due to their comparatively low capture probability \citep{Pierens08}.}. Table (\ref{t:mreseccen}) shows the eccentricity of the four giant planets, which are set to be in the same plane. On the other hand, the terrestrial planets were put in the same location as where they are currently\footnote{We note that the angular momentum deficit of the terrestrial planet system may have been somewhat lower in the past. However, this does not affect our analysis appreciably because to leading order, the frequencies of the secular system are set only by the semi-major axes and masses \citep{Murray99}.}, motivated by the analysis of \citet{Brasser09}. We ignore the impacts from scattered planetesimals, since their effects are negligible here. The duration of each integration spanned 50 Myr.

\begin{table}
\caption{Multi-resonant states.}
\label{t:mres}
  \centering
	\begin{tabular}{ |l|l|l|l| }
    \hline
       & $P_{\rm Jupiter} : P_{\rm Saturn}$ & $P_{\rm Saturn} : P_{\rm Uranus}$ & $P_{\rm Uranus} : P_{\rm Neptune}$ \\ \hline
   N1 & $2 : 3$ &   $2 : 3$ & $4 : 5$\\ 
    N2 &  $2 : 3$ &  $3 : 4$ &  $2 : 3$\\
   N3 &  $2 : 3$ & $3 : 4$ & $3 : 4$ \\ 
    N4 & $1 : 2$ & $3 : 4$ & $3 : 4$\\
    \hline
	\end{tabular}
\end{table}



The characteristic secular frequencies, obtained by Fourier analysis of the quantity $z=ie^{\imath\Omega}$, where $i$ is the inclination, $\Omega$ is the longitude of ascending node, and $\imath = \sqrt{-1}$ are shown in Figure (1). The curves corresponding to the various multi-resonant giant planet configurations are labeled accordingly. As shown in the figure, the maximum frequencies with non-negligible amplitudes are around $-18^{''}/yr$. This is similar to the current maximum large-amplitude frequency ($s_3=-18.8512^{''}/yr$) \citep{Laskar90}.

As a consequence of tidal evolution, the torque exerted on the Earth by the Moon varies as a function of time. Specifically, as the Lunar orbit expands, the spin rate of the Earth slows down and the torque becomes weaker. The tidal dissipation inside the Earth and the Moon depends on the underlying rheology and is generally complicated (see \citealt{Efroimsky08} for a review). However, as the total angular momentum remains constant under tidal dissipation, the torque and the spin precession frequency caused by the Moon can be evaluated as a function of the Earth-Moon semi-major axis ($a$). The expression for the forced spin precession frequency is $\dot{\psi} = \alpha \cos{(\varepsilon)}$, where $\psi$ is the longitude of the spin-axis, $\varepsilon$ is the obliquity, and $\alpha$ is the precession coefficient defined as \citep{deSurgy97}:
\begin{eqnarray}
\label{e:alpha}
\alpha &=& \frac{3G}{2\omega}\Big[\frac{m_{\odot}}{(a_{\oplus}\sqrt{1-e_{\oplus}^2})^3} \\ \nonumber
			&+&\frac{m_{M}}{(a_{M}\sqrt{1-e_{M}^2})^3}(1-\frac{3}{2}\sin^2i_M)\Big]E_d.
\end{eqnarray}

In the above expression, $m_{\odot}$ is the mass of the Sun, $a_{\oplus}$ and $e_{\oplus}$ are the semi major axis and the eccentricity of the Earth's orbit, $m_{M}$ is the mass of the moon, $a_{M} = a$, $e_{M}$ and $i_{M}$ are the semi major axis, eccentricity and inclination of the Moon's orbit around the Earth, $\omega$ is the spin of the Earth, and $E_d = (2C-A-B)/C$ is the dynamical ellipticity of the Earth, where $A$, $B$ and $C$ are the moment of inertia in the three principle axes. We set $E_d$ to be proportional to $\omega^2$, as it arises from rotational deformation \citep[e.g.][]{Murray99}. We plot the forced spin-axis precession rate, $\dot{\psi}$ due to both the Sun and the Moon in Figure \ref{f:alphadecay}, where the solid and the dashed curves corresponds to null ($\epsilon = 0$ deg) and nearly lateral ($\epsilon = 85$ deg) obliquities. Note that in this approach, we require the Moon to be sufficiently far away from the Earth ($a \gtrsim15 R_{\oplus}$) for the Moon's orbit to precesses about the ecliptic plane. Additionally, we over-plot the maximum orbital frequencies obtained from the N-body simulations (also shown in Figure \ref{f:FT}). 

\begin{table}
\caption{Orbital properties.}
\label{t:mreseccen}
  \centering
	\begin{tabular}{ |l|l|l|l|l|}
    \hline\hline
       & $e_{\rm Jupiter}$ & $e_{\rm Saturn}$ & $e_{\rm Uranus}$ & $e_{\rm Neptune}$ \\ \hline
   N1 & 0.0060 &   0.025 & 0.031 & 0.0083 \\ 
    N2 &  0.0038 &0.017 &  0.017 & 0.0064 \\
   N3 &  0.0069 & 0.026 & 0.016 & 0.018\\ 
    N4 & 0.044 & 0.025 & 0.053 & 0.0046 \\ \hline
           & $a_{\rm Jupiter}$ (AU) & $a_{\rm Saturn}$ (AU) & $a_{\rm Uranus}$ (AU) & $a_{\rm Neptune}$ (AU) \\ \hline
   N1 & 5.88 & 7.89 & 10.38 & 12.01 \\ 
    N2 & 5.87 & 8.00 & 9.98 & 13.16 \\
   N3 &  5.84 & 7.83 & 9.67 &11.63\\ 
    N4 & 5.48 & 8.74 & 10.59 & 12.86 \\
    \hline\hline
	\end{tabular}
\end{table}

The denoted curves suggest that for all reasonable choices of parameters, the spin-axis precession frequency has consistently exceeded the maximal secular frequencies significantly, even though the past orbital frequencies are larger than the current ones. Given that the current obliquity is relatively low, this indicates that the spin-axis resonant encounters of low order are unlikely to have played an important role in the past history of the Earth-Moon system. In other words, the EarthÕs obliquity did not vary substantially throughout the Solar System's lifetime. On the other hand, had the EarthÕs primordial obliquity been greater than $\epsilon \gtrsim 80$ deg, resonant dynamics of the spin-axis could have been possible after a few hundred Myr of tidal evolution.


Moreover, at $\sim 600$ Myr, the four giant planets reach instability and quickly scatter divergently. The onset of this transient behavior can arise from an encounter of a planet pair with a mean motion resonance (e.g. Jupiter \& Saturn's encounter with a 2:1 or a 5:3 MMR; see \citealt{Tsiganis05, Morby07, Batygin10}), or from the destruction of the resonant phase-protection mechanism by interactions with a distant self-gravitating planetesimal disk \citep{Levison11,Nesvorny12}. To mark the time when the instability occurs in Figure \ref{f:alphadecay}, the Earth-Moon distance as a function of time needs to be calculated. Assuming a constant time lag (CTL) tidal model with $t_{\rm{diss}} = 33.18$ minutes, the Earth-Moon distance at $\sim 600$ Myr is marked with an orange line in Figure \ref{f:alphadecay}. The specific choice for $t_{\rm{diss}}$ is adopted so that the Earth-Moon distance evolves to its current state at $\sim 4.5$ Gyr.

\begin{figure}[t]
\includegraphics[width=\columnwidth]{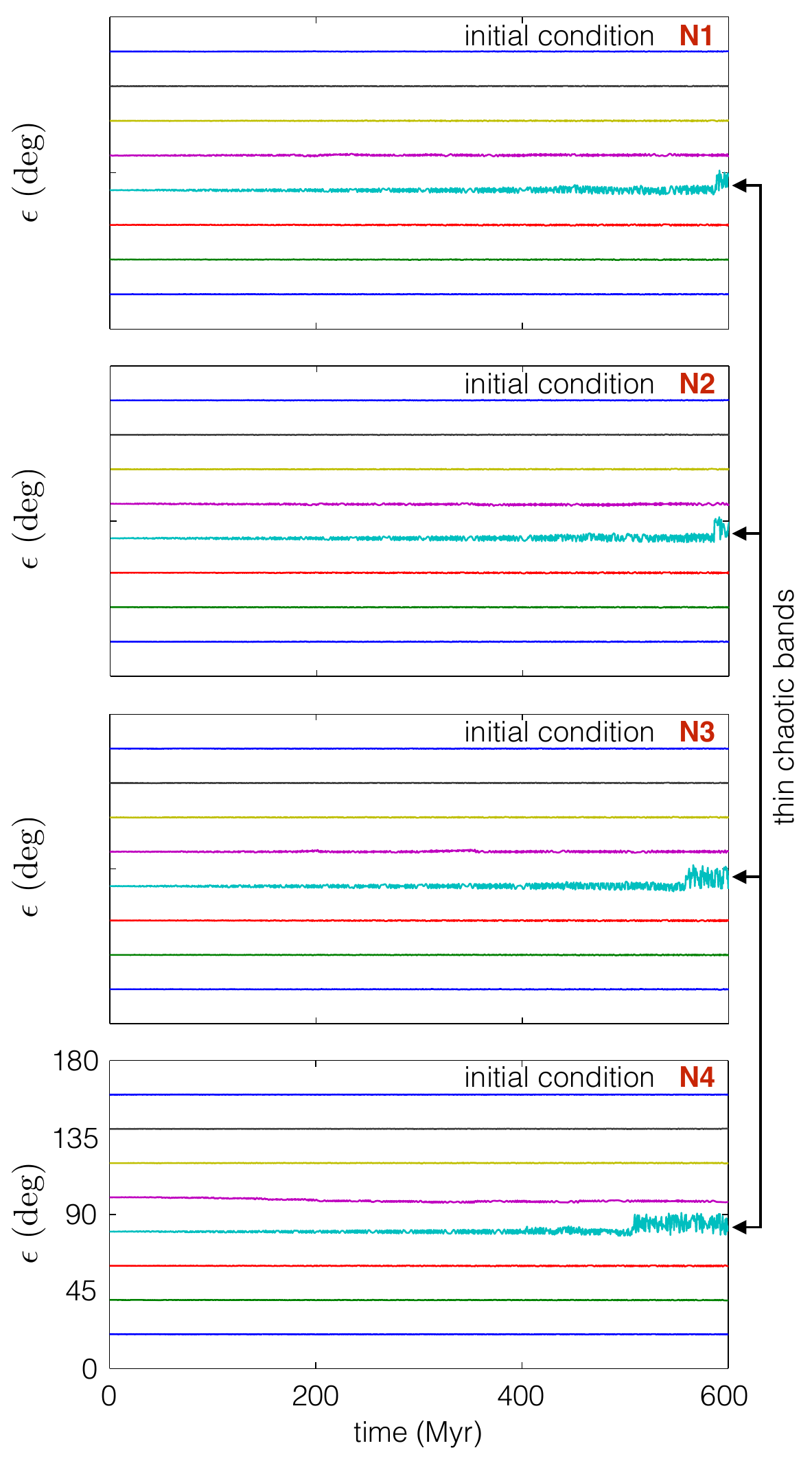}
\caption{\label{f:FR} The obliquity as a function of time when the giant planets are in resonant states, as presented in table (\ref{t:mres}). The Earth-Moon system is taken to evolve under tidal dissipation with a constant time lag ($t_{\rm{diss}} = 33.18$ minutes). The obtained solutions suggest that the obliquity remains constant except when it is initialized at $90^{\circ} \gtrsim \epsilon \gtrsim 80^{\circ}$.}
\vspace{0.1cm}
\end{figure} 

Shortly after the onset of the instability, the giant planets evolve onto their current locations with higher eccentricities and inclinations, which damp as a result of interactions with a massive planetesimal disk \citep{Levison08}. Numerical integration shows that the relevant secular frequencies when the eccentricity and inclination are damping are similar to the current frequencies, as the frequencies are largely determined by the semi major axes and masses alone \citep{Murray99}. This suggests that there are no large obliquity variations during the damping era either.

We note that an initial Solar System configuration that harbored more than two ice giants beyond the orbit of Saturn is a distinct possibility within the framework of the Nice model \citep{Nesvorny11,Batygin12,Nesvorny12}. Although such configurations will yield quantitatively different evolutions from the cases presented here, the synonymity of the Fourier decompositions of the initial conditions presented in this work (see Figure \ref{f:FT}) suggest that the introduction of additional Neptune-mass planets into the resonant chains is unlikely to alter the results significantly. Accordingly, here we use only the four giant planet models as illustrative examples.

\section{Numerical Integrations}
The analysis performed above suggests that the obliquity does not exhibit large variations due to the resonances as a result of Chirikov diffusion arising from secular spin-orbit resonance overlap at low obliquities. This however does not negate the possibility of substantial obliquity diffusion associated with stochastic pumping arising from a chaotic orbit \citep{Lichtenberg83}. To illustrate explicitly the evolution of the obliquity as a function of time, we numerically integrate the obliquity variation using the Earth's orbital evolution obtained from N-body simulations. Specifically, we evolve the Earth's obliquity adopting multi-resonant conditions for the giant planets (depicted in Table 1), and taking the precession coefficient to change in accord with the Moon's tidal recession and the Earth's spin-down. As before, a CTL tidal model was employed with $t_{\rm{diss}} = 33.18$ minutes. 

The Hamiltonian describing the evolution of the obliquity is well documented in the literature \citep[e.g.][]{Colombo66, Laskar93a, Touma93, deSurgy97}:
\begin{eqnarray}
\label{eqn:HF}
H(\chi, \psi, t) &=& \frac{1}{2}\alpha\chi^2 +\sqrt{1-\chi^2} \\ \nonumber
						  &\times& (A(t)\sin{\psi}+B(t)\cos{\psi})),
\end{eqnarray}
where $\chi = \cos \varepsilon$, $\alpha$ is the precession coefficient (see eqn \ref{e:alpha}), and 
\begin{eqnarray}
A(t) = 2(\dot{q}+p(q\dot{p}-p\dot{q}))/\sqrt{1-p^2-q^2}, \\
B(t) = 2(\dot{p}-q(q\dot{p}-p\dot{q}))/\sqrt{1-p^2-q^2},
\end{eqnarray}
where $p=\sin{i_{\oplus}/2}~\sin{\Omega_{\oplus}}$ and $q=\sin{i_{\oplus}/2}~\cos{\Omega_{\oplus}}$. The integrations are taken to span 600 Myr.


Starting with different initial obliquities, we plot the numerical results in Figure \ref{f:FR}. The calculations suggest that independent of the initial condition, when $\varepsilon \lesssim 80^{\circ}$, the obliquity remains nearly constant as a function of time. The obliquity varies substantially and rapidly when $90^{\circ} \gtrsim \varepsilon \gtrsim 80^{\circ}$, because the precession frequency ($\alpha \cos{\varepsilon}$) matches with the orbital perturbation frequency at later times (as shown in Figure \ref{f:alphadecay}). This is consistent with our analysis of the resonances, indicating the effect of chaotic pumping on the obliquity diffusion is negligible here. As noted above, this suggests that that the Earth's present obliquity has been preserved throughout the system's history.

\section{Summary}
It is generally accepted that substantial modulation of the EarthÕs obliquity can result in dynamically-forced climatological changes. In turn, the variation of a planet's obliquity is sensitive to the precession frequency of its spin axis and its secular orbital frequencies. When the two sets of frequencies match, resonances may arise and the planet's obliquity may undergo large amplitude variations (as the case for the Moonless Earth \citep[e.g.][]{Laskar93a, Li14} or for Mars \citep{Ward73, Touma93, Laskar93b}). Currently, the Earth's obliquity is regular and oscillates with a very small amplitude ($22.1-24.5^{\circ}$). However, the orbital forcing of the Earth likely underwent significant changes throughout the Solar SystemÕs dramatic history, potentially suggesting that the dynamical state of the EarthÕs spin-axis may have been resonant in the past. In this study, we have quantified this possibility.

We began our investigation by examining the feasible proximity of the pre-instability secular modes that characterize the orbital evolution of the Earth to the precession rate of the EarthÕs spin-axis. To obtain the secular orbital frequencies of the Earth, we computed the orbital evolution of the Earth, adopting Solar System orbital architectures that were previously demonstrated to serve well as initial conditions for the Nice model by \citet{Batygin10}. Our analysis has shown that even under favorable assumptions, the slow-down in the spin-axis precession frequency associated with the tidal evolution of the Earth-Moon system as well as the increase in the forcing frequencies associated with a more compact giant-planet configuration are insufficient to give rise to secular spin-orbit resonant encounters in the system at low obliquities. 

Subsequently, to illustrate the obliquity variation explicitly, we directly integrated the pre-LHB spin-axis evolution of the Earth. We adopted the Earth's orbital evolution from the N-body simulations and calculated the precession coefficient based on the CTL tidal dissipation model for the Earth-Moon system. The numerical results show that only minimal oscillations in the EarthÕs obliquity can be expected for primordial obliquities less than $\epsilon \lesssim 80$ deg. Indeed, this is consistent with our qualitative analysis of spin-axis resonant conditions. 

Cumulatively, our study shows that the dynamical perturbations arising from the other planets are unlikely to have given rise to resonant excitations of the EarthÕs spin-axis at low obliquity. Moreover, chaotic pumping arising from a diffusing orbit also leads to negligible evolution. Thus, the Earth's obliquity likely did not vary substantially throughout the dramatic lifetime of the Solar System, and was probably set in situ by the giant impact associated with the formation of the Moon. This remarkable aspect of Solar System dynamics renders the EarthÕs obliquity one of the few truly primordial features of the Solar System.

\acknowledgments

\bibliographystyle{hapj}


\end{document}